# The Gain in RS

I.V. Dovgan*
Department of Physics, Moscow State Pedagogical University, Moscow 119992, Russia.
*dovganirv@gmail.com

**Abstract:** The calculated linear gain in relativistic strophotron is averaged over electron distribution. It is shown, that after averaging resonance peaks remain in amplification spectrum. The nonresonant background gives only negligibly small contribution to the gain.

1. **Introduction**

In papers [1-45] Free electron lasers based on undulators and strophotrons are investigated. In [3] the emitted energy in RS was calculated. The dependence of resonance frequency on transverse coordinate in RS brings to strong inhomogeneously broadening of amplification spectrum and the lines overlap.

The calculated emitted energy in the relativistic strophotron [3] should be averaged over electron distribution. In the present article this averaging is presented.

2. **The Averaged Gain**

The expression (10) of [3] for the gain at the main frequency $\omega_{res}$, is correct under the conditions $K < 1$ and $d_e < \Delta x_0$ where $\Delta x_0$ is determined by the right-hand side of inequality (6) of [3]. Otherwise, under the condition (6) ($d_e > \Delta x_0$), the gain must be averaged over $x_0$, and this procedure can essentially changes the result of [3].

For the beam as a whole, under the condition (6) of [3], again there is a strong inhomogeneous broadening. It means that in a general case, after averaging over $x_0$ many terms of the sum over s give their contribution to the energy $\Delta\varepsilon$ emitted at a given frequency $\omega$. The nonresonance part of the averaged emitted energy is found similarly $\left(\dfrac{\overline{d\mathrm{E}\omega}}{d\omega do}\right)_{n.r.}$ (15) of [3] and has the form

$$\overline{\Delta\varepsilon}_{n.r.} = \frac{\sqrt{3}Te^2E_0^2\omega^{1/2}}{32\sqrt{2^{1/2}}\varepsilon_0\tilde{d}_e\pi^{5/2}} \sum_{s-s_{min}\geq 1} (s-s_{min})^{-3/2} \left\{\varphi\left(x_0^{(s)}\right) - x_0^{(s)}\varphi\left(x_0^{(s)}\right)\right\} \qquad (1)$$

where $x_0^{(s)}$, $s_{min}$ and $\varphi\left(x_0^{(s)}\right)$ are given by:

$$\left(x_0^{(s)}\right)^2 = \frac{4}{\Omega\omega}(2s+1) - \frac{\alpha^2}{\Omega^2} - \frac{2}{\gamma^2\Omega^2} \equiv \frac{8}{\Omega\omega}(s-s_{min}) \geq 0$$

$$s_{min} = \frac{\alpha^2\omega}{8\Omega} + \frac{\omega}{4\gamma^2\Omega^2} - \frac{1}{2}$$

$$\varphi\left(x_0^{(s)}\right) = \exp\left(-\frac{x_0^2}{\tilde{d}_e^2}\right)\left(\alpha^2 + x_0^2\Omega^2\right) \qquad (2)$$

$$\times \left(\frac{1}{2\gamma^2} + \frac{\alpha^2 + x_0^2\Omega^2}{4}\right)\left[J_s(Z(x_0)) - J_{s+1}(Z(x_0))\right]^2.$$

The sum over s (1) can be replaced by the integral over $x = \left(8/\Omega\omega\tilde{d}_e^2\right)(s-s_{min})$ Under the conditions $\alpha\gamma \gg 1$, $\omega \gg 2\gamma^2\Omega$, and if $\omega < \omega_{max}$, the result is given by

$$\overline{\Delta\varepsilon}_{n.r.} = \frac{3^{1/3}\Gamma^2\left(\frac{1}{3}\right)}{16\cdot 2^{1/6}\pi^{3/2}} \frac{Te^2E_0^2\omega^{1/2}}{\varepsilon_0\tilde{d}_e\Omega^{7/6}\omega^{5/6}}$$

$$\cdot \frac{\Delta x_0}{\tilde{d}_e} \int_{x_{min}}^{\infty} \frac{dx}{x^{3/2}}\left(\tilde{\varphi}(x) - 2x\tilde{\varphi}'(x)\right) \qquad (3)$$

where

$$x_{min} = \left(\Delta x_0/\tilde{d}_e\right)^2 \ll 1$$

and

$$\tilde{\varphi}(x) = e^{-x}\left(1 + \frac{\Omega^2\tilde{d}_e^2}{\alpha^2}x\right)^{2/3}. \qquad (4)$$

Now the lower limit of integration in (3) may not be replaced by 0 because the integrand has too strong a singularity at the point x = 0. These are small x around $x_{min}$ which give the main contribution to the integral (3). For this reason, the nonresonance emitted energy (3) can be estimated as

$$\overline{\Delta\varepsilon}_{n.r.} = \frac{Te^2E_0^2\alpha^{4/3}}{\varepsilon_0d_e\Omega^{7/6}\omega^{5/6}} \qquad (5)$$

This is an estimate of the nonresonant background in an amplification of the external wave.

Again the resonances in the emitted energy $\overline{\Delta\varepsilon}$ appear when $s_{min}$ (5) is close to some integer $|s_{min} - N| \ll 1$, $N = [s_{min}]$ or $N = [s_{min}] + 1$. A physical nature of these resonances is the same as previously: they correspond to the threshold of appearance of a new channel of amplification. The threshold corresponds to $x_0 \approx 0$. The factor $(d/du_s)(\sin^2 u_s / u_s^2)$ in the sums (1) has a stronger singularity at the threshold than $\delta'(x_0)$. This is why this threshold term being averaged over $x_0$ gives an anomalously large contribution to $\overline{\Delta\varepsilon}$. The result has the form

$$\overline{\Delta\varepsilon}_{res} = \frac{T^{5/2} e^2 E_0^2 \alpha^{1/2}}{3\pi^{5/2} 8\sqrt{2}\varepsilon_0 \tilde{d}_e \Omega \gamma^4} K_{2/3}^2\left(\frac{\omega_N}{\omega_{max}}\right) \tilde{F}_{res}(\omega) \tag{6}$$

where again $\omega_N = (4\Omega/\alpha^2)(2N+1)$, $\omega_{max} = 3\gamma^3 \alpha \Omega$. The factor of the resonance curve $\tilde{F}_{res}(\omega)$ is given by

$$\tilde{F}_{res}(\omega) = \int \frac{d\xi}{\xi} \frac{d}{d\xi} \frac{\sin^2\left[\xi^2 + \frac{\alpha^2 T}{8}(\omega - \omega_N)\right]}{\left[\xi^2 + \frac{\alpha^2 T}{8}(\omega - \omega_N)\right]^2} = \frac{\alpha^2 T}{8} \frac{d}{d\omega} F_{res}(\omega) \tag{7}$$

where $F_{res}(\omega) = \int d\xi \frac{\sin^2\left[\xi^2 + \frac{\alpha^2 T}{8}(\omega - \omega_N)\right]}{\left[\xi^2 + \frac{\alpha^2 T}{8}(\omega - \omega_N)\right]^2}$ is the form factor of the resonance curve in a spontaneous emission.

If $\omega < \omega_{max}$, with the aid of the approximation for McDonalds function

$K_{2/3}(u) \approx \frac{\Gamma(2/3)}{2^{1/3} u^{2/3}}$, $|u| \ll 1$, (6) is reduced to

$$\overline{\Delta\varepsilon}_{n.r.} = \frac{3^{1/3} \Gamma^2\left(\frac{2}{3}\right)}{16 \cdot 2^{1/6} \pi^{5/2}} \frac{T^{5/2} e^2 E_0^2 \Omega^{1/3} \alpha^{4/3}}{\varepsilon_0 \tilde{d}_e \omega_N^{5/6}} \tilde{F}_{res}(\omega). \tag{8}$$

Comparison with $\overline{\Delta\varepsilon}_{n.r.}$ (5) gives an estimate

$$\overline{\Delta\varepsilon}_{res} / \overline{\Delta\varepsilon}_{n.r.} = (\Omega T)^{3/2} \gg 1. \tag{9}$$

In a stimulated emission, the relation between the height of the resonance peaks and the nonresonant backgrounisd inverse as compared to a spontaneous emission. The nonresonant background (5) gives only a negligibly small contribution to the gain which is determined mainly by the resonance emitted energy $\overline{\Delta\varepsilon}_{res}$ (6), (8):

$$G = \eta \frac{8\pi N_e}{E_0^2} \Delta \varepsilon_{res} = \frac{\eta(\ln 2)^{1/2}}{\pi^{3/2}} \frac{N_e \Gamma_0 L^{5/2}}{\sqrt{d_e} \gamma^{7/2}} \left(\frac{\omega}{\omega_{max}}\right)^{1/2} K_{2/3}^2 \left(\frac{\omega}{\omega_{max}}\right) \tilde{F}_{res}(\omega) \qquad (10)$$

where $\alpha = d_e \Omega$, $N_e$ is the electron number density and $\eta$ is the artificially introduceed factor taking into account the difference of dimensions of the beam outside and inside the strophotron. This difference appears due to the electron oscillations which effectively broaden the beam. If we take $(x_0)_{max} \approx d_e/2$, the maximum amplitude of oscillations is $a_{max} = ((1/4)d_e^2 + (\alpha^2/\Omega^2))^{1/2}$ and

$$\eta = \frac{d_e/2}{a_{max}} = \left(1 + \frac{4\alpha^2}{d_e^2 \Omega^2}\right)^{-1/2} \qquad (11)$$

If, in particular, $d_e = \alpha/\Omega$, $\eta = 1/\sqrt{5}$.

### 3. Conclusion

The linear gain is calculated and averaged over electron distribution in relativistic strophotron. It is shown that after averaging resonance peaks remain in amplification spectrum. The nonresonant background gives only negligibly small contribution to the gain. The resonance peaks are higher than nonresonant background as far as $(\Omega T)^{3/2} \gg 1$, i.e. the electron transverse oscillation number is higher.